\documentclass[fleqn,12pt,twoside]{article}
\usepackage{espcrc1}

\usepackage{graphicx}
\usepackage[figuresright]{rotating}

\newcommand{\AmS}{{\protect\the\textfont2
  A\kern-.1667em\lower.5ex\hbox{M}\kern-.125emS}}

\title{Photoproduction of the doubly-strange $\Xi$ Hyperons}

\author{J.W. Price\address{Department of Physics and Astronomy, \\ 
        University of California at Los Angeles, Los Angeles, CA 90095
        USA}, %
        J. Ducote\addressmark,
        J. Goetz\addressmark,
        and
        B.M.K. Nefkens\addressmark, for the CLAS Collaboration}
       
\begin{document}

\maketitle

\begin{abstract}
We report on the first measurement of exclusive $\Xi^-$ and $\Xi^0$
photoproduction.  The $\Xi^-$ states are produced in the reaction $\gamma
p\to K^{+}K^{+}\Xi^-$, and the $\Xi^0$ states in $\gamma p\to
K^+K^+\pi^-\Xi^0$.  Identification is made by the unique mass measured
as the missing mass of the $K^{+}K^{+}$ (or $K^+K^+\pi^-$) system
using the CLAS detector at the Thomas Jefferson National Accelerator
Facility.  A systematic study of the excited $\Xi$ spectrum improves
our understanding of the $N^*$ and $\Delta^*$ states, since the
$\Xi^*$ states are related to them by $SU(3)$ flavor symmetry.  At the
highest energies available at Jefferson Lab, we begin to find evidence
for known excited $\Xi^-$ states in the photoproduction process, and
possibly new states at 1770 and 1860 MeV, although we do not have
enough statistics to draw a strong conclusion.  A search for the
$\Xi^{--}_5(1862)$ pentaquark state seen by NA49 is made using the
process $\gamma p\to K^+K^+\pi^+X$, but the result is inconclusive for
lack of statistics.
\end{abstract}

\section{Introduction}

We know very little about the doubly-strange $\Xi$, or cascade,
hyperons.  Although $SU(3)_F$ symmetry predicts the existence of a
$\Xi$ for every nucleon \emph{and} a $\Xi$ for every $\Delta$, for a
total of 44 $\Xi$ states~\cite{Nef95}, only eleven have been seen to
date~\cite{PDG02}.  Of these eleven, only three have been completely
identified by their mass, width, spin, and parity. Less still is known
about their production mechanisms and decay branching ratios.  

The bulk of our knowledge on the cascade spectrum has come from kaon
beams, with some information from hyperon beams.  It is important to
find a new means of producing the $\Xi$, since there is currently no
suitable kaon facility.  Ref.~\cite{Nef96} first suggested using the
photoproduction process $\gamma p\to K^+K^+\Xi^-$ to look for the
cascade.  The threshold for the production of the ground state
$\Xi^-(1321)$ using this process is 2.37 GeV.  For the $\Xi^0$, an
extra $\pi^-$ is detected, and the threshold for the production of the
ground state is 2.73 GeV.  These processes provide unique event
signatures, in which two $K^+$'s are required.  Their exclusive nature
results in very little physics background.  The Thomas Jefferson
National Accelerator Facility (JLab), with its 5.7 GeV tagged photon
beam~\cite{Sob00} and the CLAS detector~\cite{Mec03} is an excellent
place for this study. 

This study is motivated by the observation that the $\Xi$ states are
approximately nine times narrower than the nucleon or $\Delta$
states~\cite{Cox00}.  This was first explained by Riska~\cite{Ris01}
to be related to the number of light quarks within the baryon.  The
narrower $\Xi$ states are much easier to detect than the $N^*$ or
$\Delta^*$ states, and are expected to be visible in a simple missing
mass spectrum.

\section{Data}

There has as yet been no dedicated JLab experiment to search for
cascade states.  However, there are three existing CLAS data sets
(designated $g6a$, $g6b$, and $g6c$) taken for other purposes that are
compatible with $\Xi$ photoproduction.  The relevant running
conditions for each of these data sets are given in
Table~\ref{tab:runs}. 
\begin{table}[htb]
\caption{The three CLAS data sets used for $\Xi$ photoproduction.  The
  different columns show the tagged photon energy range $E_\gamma$,
  the integrated luminosity$\int {\cal L}dt$, the position of the
  target $z_{tgt}$ (0 is at the center of CLAS), and the relative CLAS
  torus current $I_t$.}  
\label{tab:runs}
\renewcommand{\tabcolsep}{2pc} 
\renewcommand{\arraystretch}{1.2} 
\begin{tabular}{@{}ccccc}
\hline
Run & 
$E_\gamma$ (GeV) & 
$\int {\cal L}dt$ ($pb^{-1}$) & 
$z_{tgt}$ (cm) & 
$I_t$ \\
\hline
$g6a$ & $3.2-3.9$ & 1.1            & $0$    & $I_0$ \\
$g6b$ & $3.0-5.2$ & not well-known & $0$    & $I_0$ \\
$g6c$ & $4.8-5.4$ & 2.7            & $-100$ & $0.58I_0$ \\
\hline
\end{tabular}
\end{table}

\section{Results}

\subsection{$\Xi^-$ photoproduction}
\begin{figure}[htb]
\begin{minipage}[t]{75mm}
\includegraphics[width=75mm]{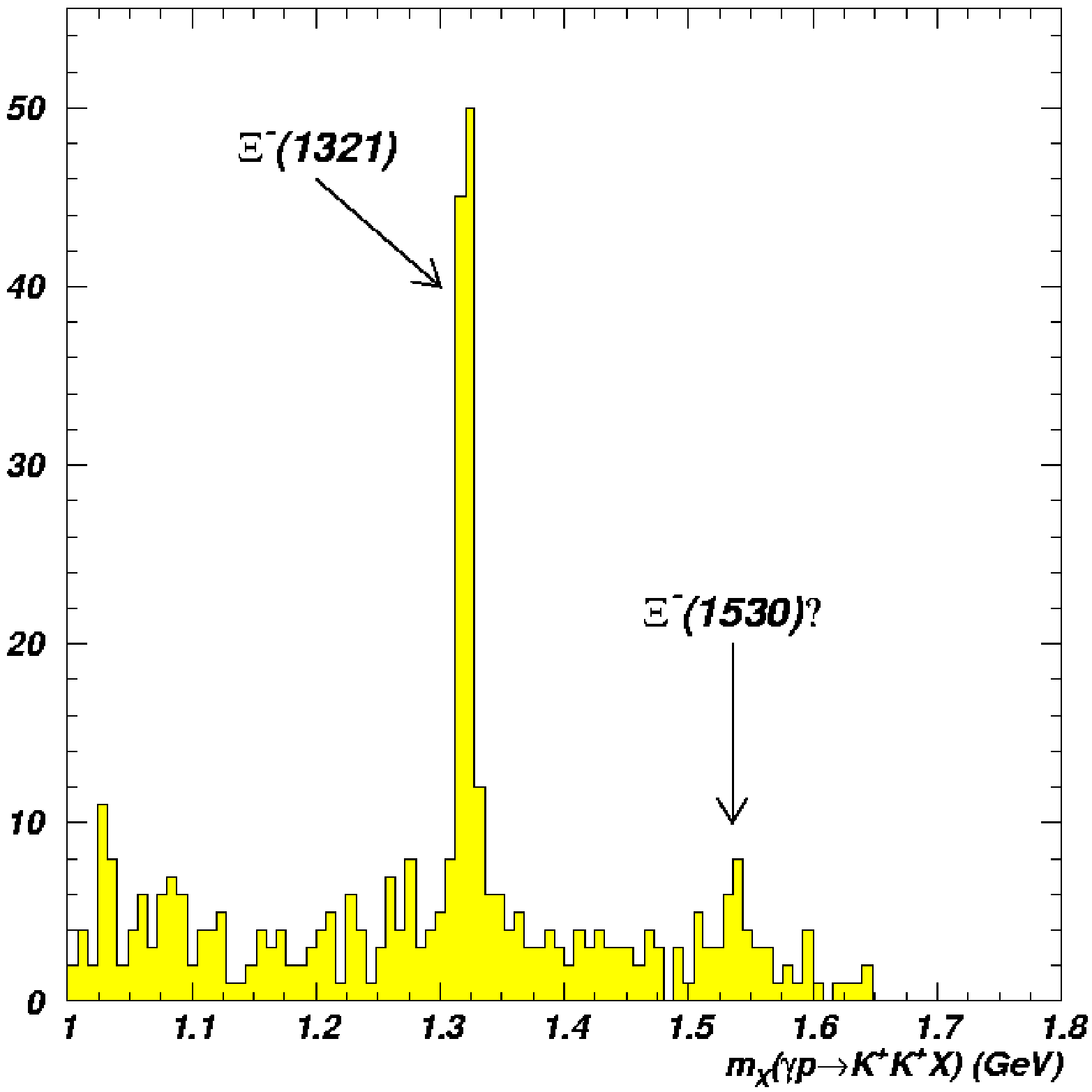}
\caption{\label{fig:g6a}The $K^+K^+$ missing mass for the $g6a$ data
  set with a tight $K^+$ particle ID.  The signal-to-noise ratio for
  the ground state $\Xi^-(1321)$ is approximately 10:1.  There is a hint
  of structure at the position of the first excited cascade state at
  1530 MeV.}
\end{minipage}
\hspace{\fill}
\begin{minipage}[t]{75mm}
\includegraphics[width=75mm]{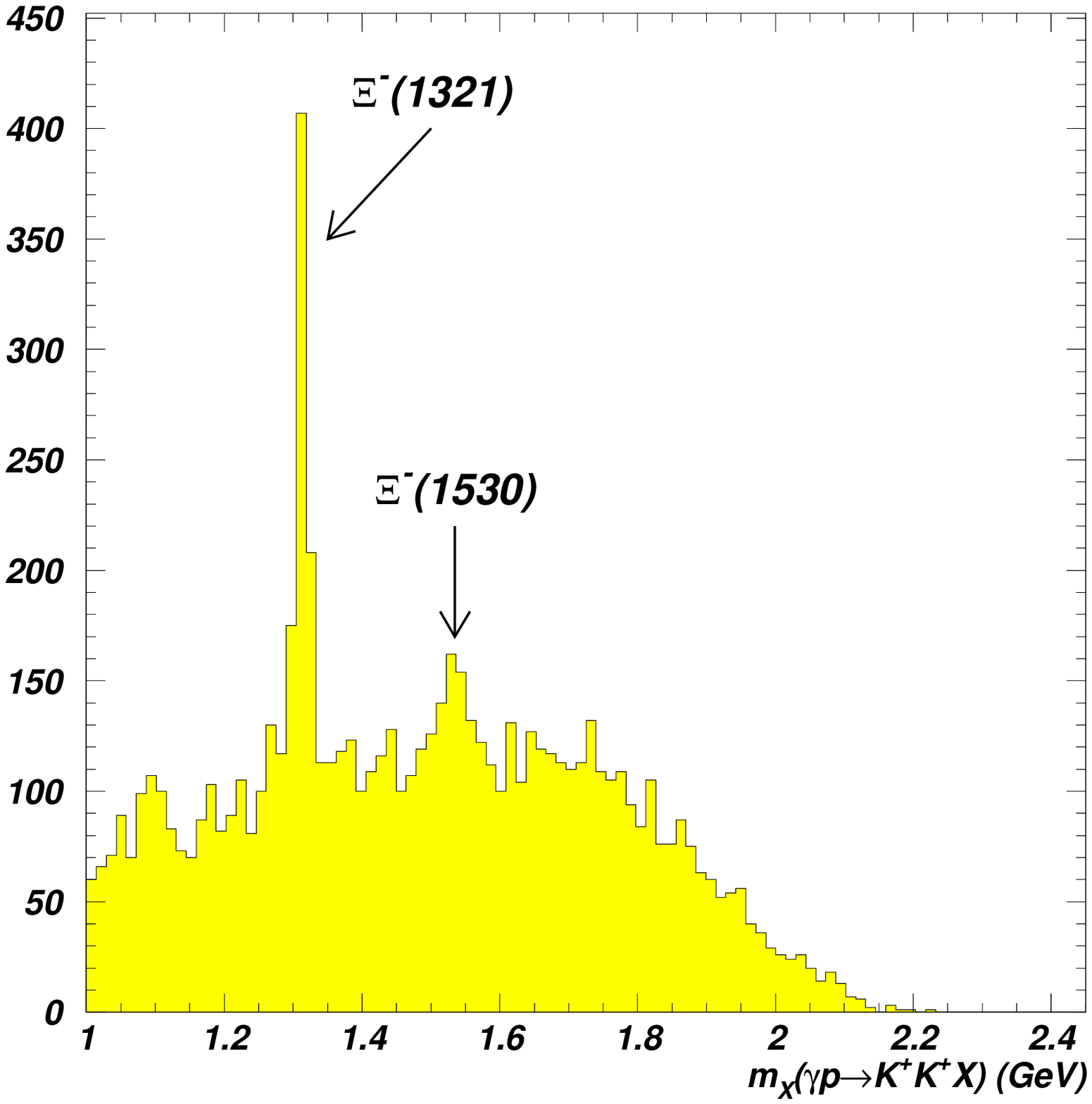}
\caption{\label{fig:g6b}The $K^+K^+$ missing mass for the $g6b$ data
  set, with loose $K^+$ particle ID.  Both the ground state and the
  first excited state are clearly seen in the data.  There is
  additional structure at 1100 MeV due to the $K^+\pi^+\Sigma^-$ final
  state in which the $\pi^+$ is misidentified as a $K^+$.}
\end{minipage}
\end{figure}

Figure~\ref{fig:g6a} shows the missing mass of the $K^+K^+$ system for
the $g6a$ data set.
The ground state cascade at 1321 MeV is clearly seen in the spectrum,
showing that there is indeed little physics background.  By using a
tight $K^+$ particle ID, we also see possible structure at the
position of the first excited state at 1530 MeV.  However, the phase
space for the production of the $\Xi^-(1530)$ is too small to show a
convincing peak under the $g6a$ running conditions.

Figure~\ref{fig:g6b} shows the same spectrum from the $g6b$ data set.
These data were taken at higher energy; both the ground state and the
first excited state of the $\Xi^-$ are clearly seen.  The $K^+$
particle ID is looser than in the $g6a$ data set, which leads to a
background due to $\pi/K$ misidentification.  The structure at 1100
MeV in Fig.~\ref{fig:g6b} is due to the process $\gamma p\to
K^+\pi^+\Sigma^-$, in which the $\pi^+$ is misidentified as a $K^+$.
This can be seen by plotting the $K^+K^+$ missing mass vs.\ the
missing mass obtained in the $\gamma p\to K^+\pi^+X$ process, by
forcing one of the kaon masses to be that of the $\pi^+$.  Improving
the particle ID for this data set is currently under study. 

\begin{figure}[htb]
\begin{minipage}[t]{75mm}
\includegraphics[width=75mm]{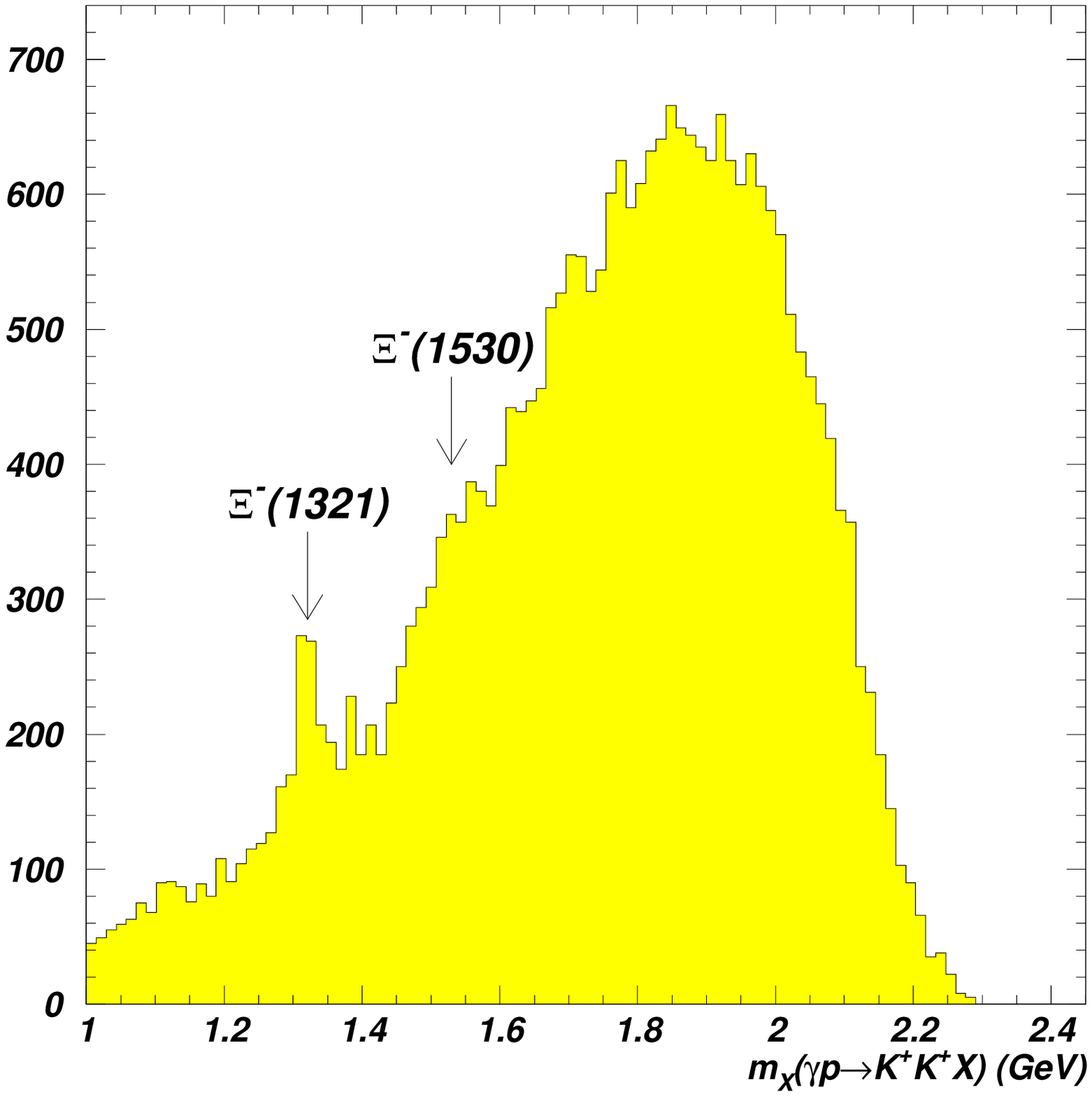}
\caption{\label{fig:g6c-init}The $K^+K^+$ missing mass for the $g6c$
  data set.  These data were taken with a very high photon flux, which
  resulted in a large background due to beam accidentals and $\pi/K$
  misidentification.  The ground state is still clearly seen, but the
  $\Xi^-(1530)$ appears only as a shoulder in the plot.}
\end{minipage}
\hspace{\fill}
\begin{minipage}[t]{75mm}
\includegraphics[width=75mm]{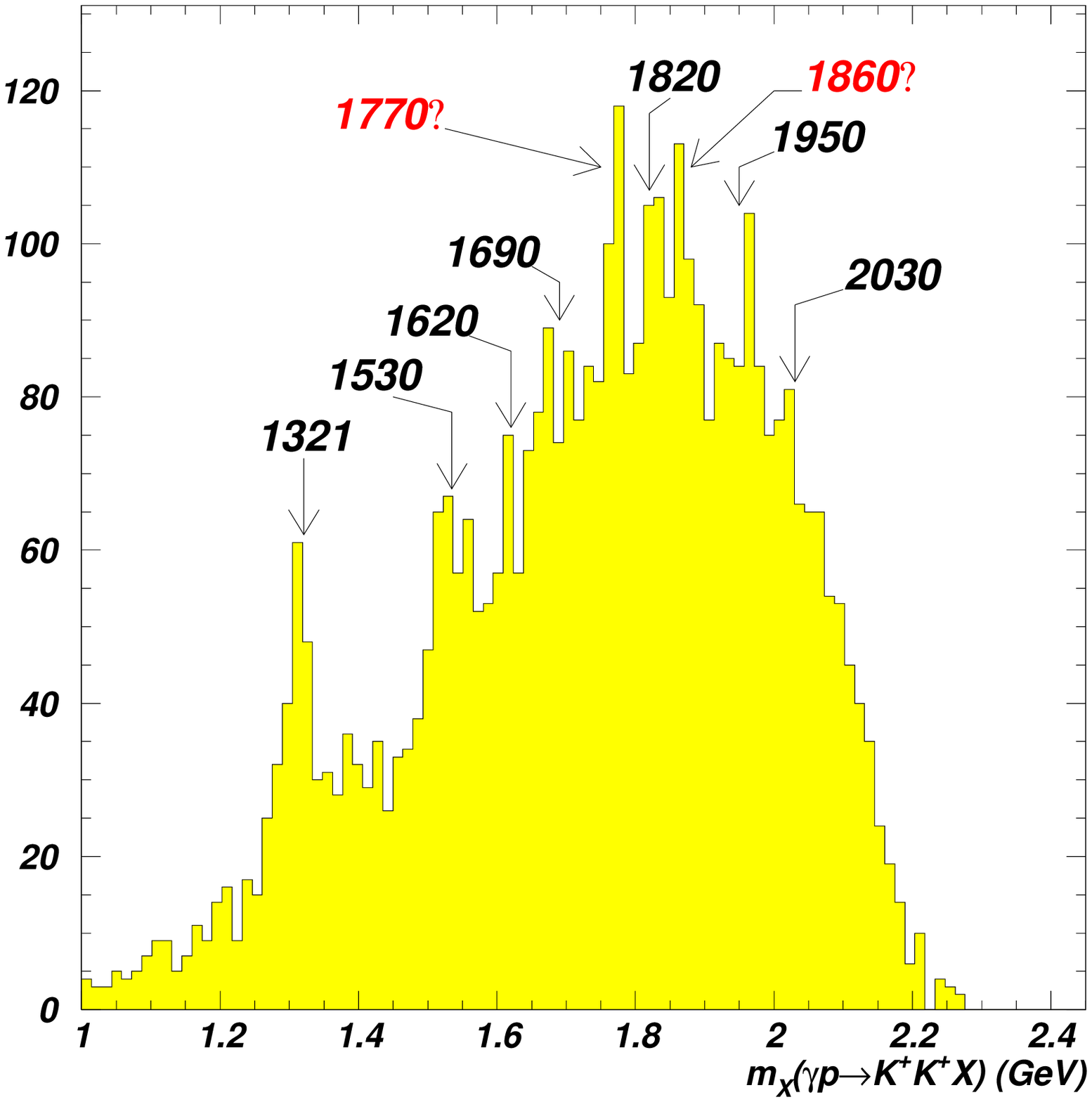}
\caption{\label{fig:g6c-final}The same plot as in
  Fig.~\ref{fig:g6c-init}, with the additional requirement of a proton
  in the final state.  This removes the backgrounds due to $\pi/K$
  misidentification, and reveals a great deal of structure at the
  higher masses.  Arrows on the plot indicate the masses of every
  cascade hyperon in the Particle Data Book.  Two structures appear in
  this plot that are not listed in the Particle Data Book, at 1770 and
  1860 MeV.}
\end{minipage}
\end{figure}

Figure~\ref{fig:g6c-init} shows the $K^+K^+$ missing mass spectrum for
the $g6c$ data set.  
The $g6c$ data set was taken at a much higher photon flux than the
$g6a$ and $g6b$ data sets.  This led to a large background due to beam
accidentals, and increased the background due to $\pi/K$
misidentification.  

The latter can be removed by considering the $\Xi^-$ decay chain
$\Xi^-\to\pi^-\Lambda\to\pi^-\pi^-p$ (or $\to\pi^-\pi^0n$).  In the
case of a single $\pi/K$ misidentification, the final state is
$K^+\pi^+\Sigma^-$; for double $\pi/K$ misidentification, it is
$\pi^+\pi^+\Delta^-$.  Neither of these processes will result in a
proton in the final state; both the $\Sigma^-$ and the $\Delta^-$
decay nearly 100\% to $\pi^-n$.  Consequently, we can remove both of
these backgrounds by requiring the presence of a proton in the final
state.  

Figure~\ref{fig:g6c-final} shows the result of this additional
requirement.  The figure shows the $K^+K^+$ missing mass for the
events in which a proton is also detected.  The $\Xi^-(1530)$ is again
clearly seen.  This plot reveals a great deal of apparent structure
not visible in Fig.~\ref{fig:g6c-init}.  To study the significance of
the enhancements in Fig.~\ref{fig:g6c-final}, we compare their
positions in the plot to known states in the Particle Data
Book~\cite{PDG02}.  These states are shown by arrows in the plot.  We
find that every state listed in the Particle Data Book up to 2030 MeV
is well-matched to an enhancement in Fig.~\ref{fig:g6c-final}
(although the evidence for the $\Xi^-(1620)$ and $\Xi^-(1690)$ in
Fig.~\ref{fig:g6c-final} is weak).  Furthermore, we note that
Fig.~\ref{fig:g6c-final} has two additional enhancements that are not
in the Particle Data Book, at 1770 and 1860 MeV.  In this mass region,
these values should be accurate to approximately 15 MeV.  Both of
these structures are robust, in the sense that they do not appear to
be an effect of the histogram binning.  We are currently studying the
possibility of reducing the background further to enhance the
structure. 

\begin{figure}[htb]
\begin{minipage}[t]{75mm}
\includegraphics[width=75mm]{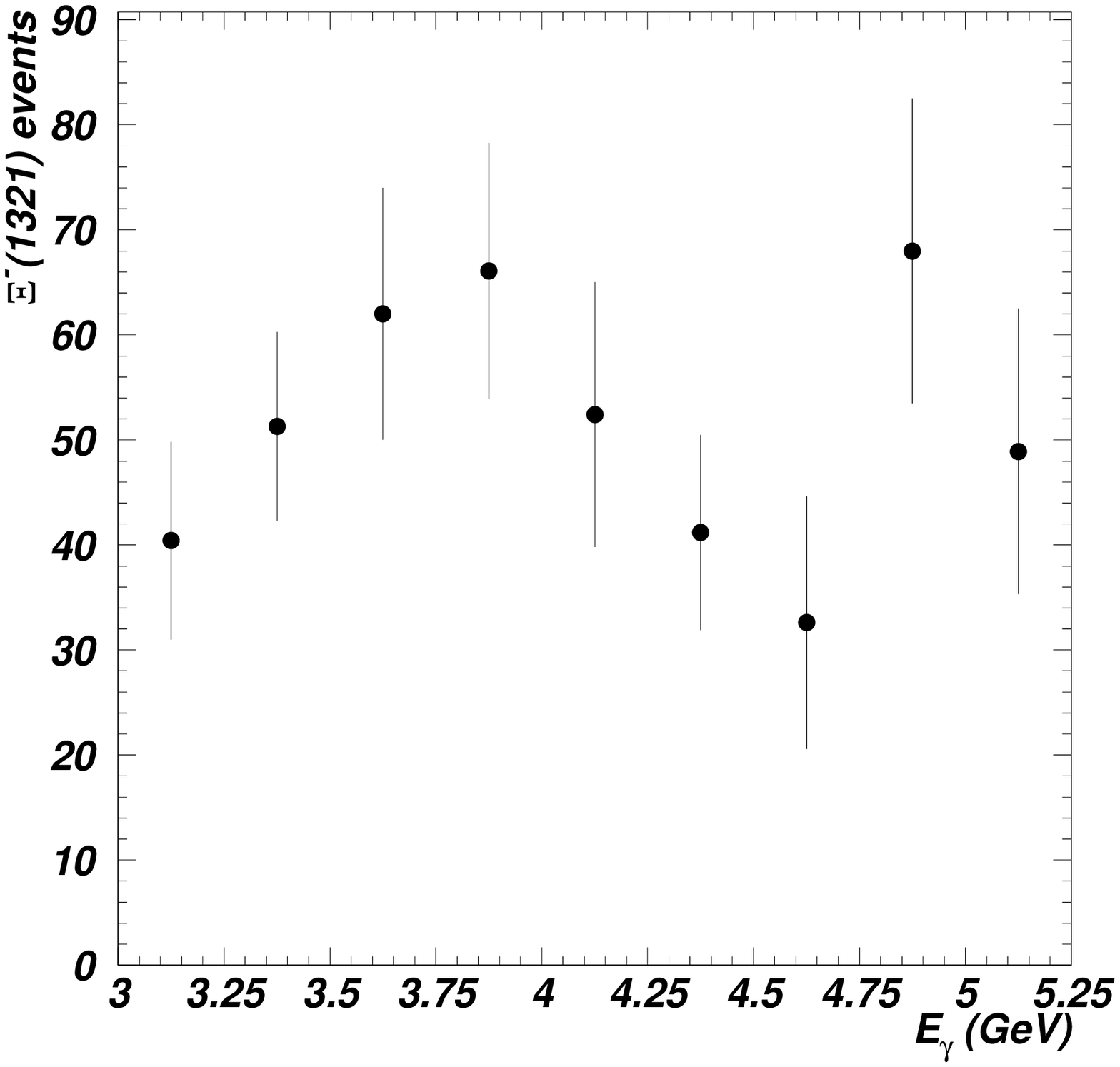}
\caption{\label{fig:yield}The total yield for $\gamma p\to
  K^+K^+\Xi^-$ as a function of the photon energy $E_\gamma$ for the
  $g6b$ data set.  The points reflect the actual number of detected
  events in each 250 MeV bin.  Neither the photon normalization nor
  the detector acceptance correction have been applied.}
\end{minipage}
\hspace{\fill}
\begin{minipage}[t]{75mm}
\includegraphics[width=75mm]{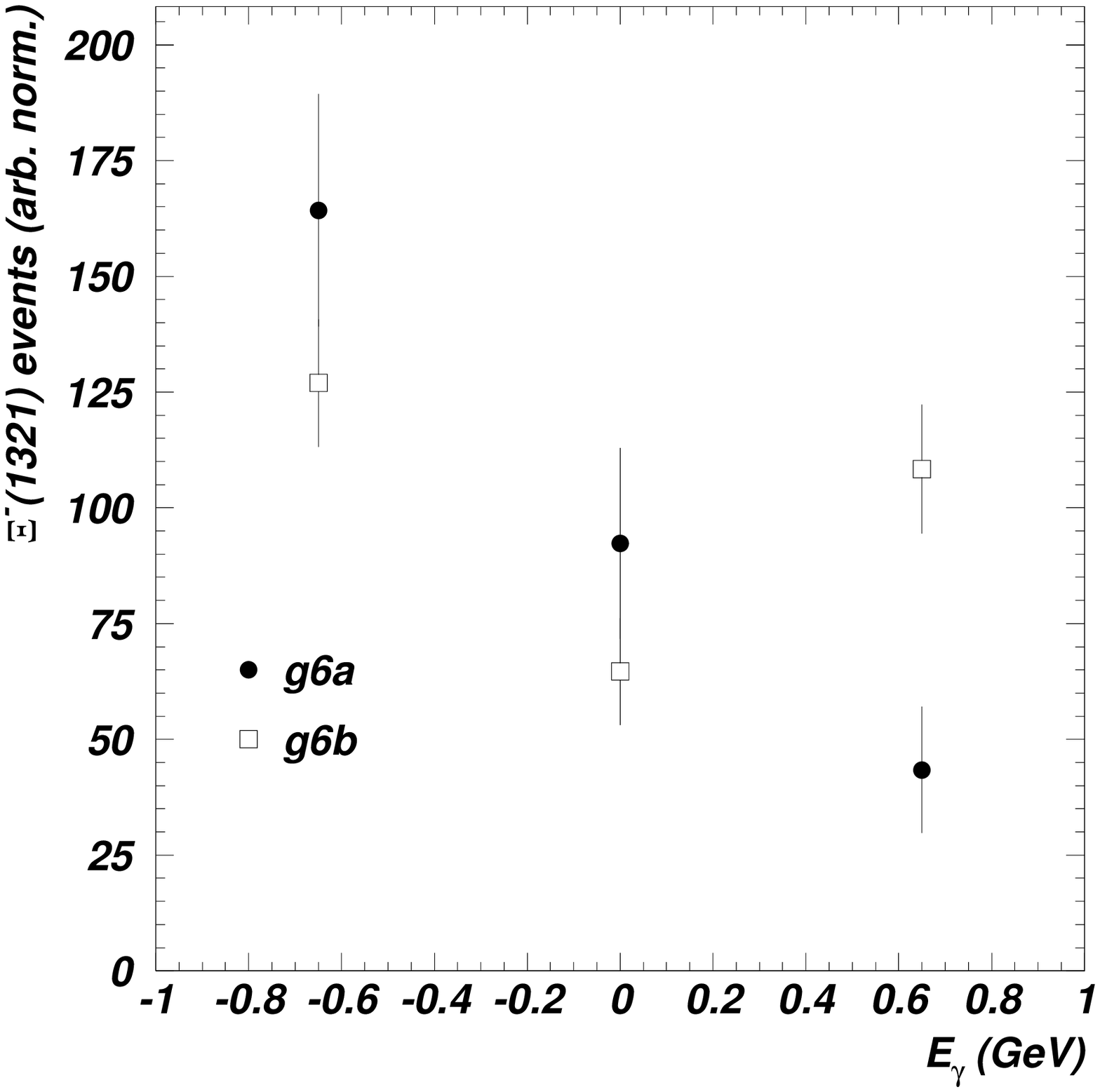}
\caption{\label{fig:angdist}The angular distribution for the $g6a$
  (squares) and $g6b$ (circles) data sets.  The event numbers are
  arbitrarily normalized.  Neither the photon normalization nor the
  detector acceptance correction have been applied.}
\end{minipage}
\end{figure}

Information on the production mechanism will be obtained by looking at
the energy dependence and angular distribution of cascade
photoproduction. These studies are planned for the near future.  The
preliminary results of this study, which are neither normalized for
the photon flux nor corrected for the detector acceptance, are shown
in Fig.~\ref{fig:yield} for the energy dependence, and in
Fig.~\ref{fig:angdist} for the angular distribution.

The preliminary nature of Figs.~\ref{fig:yield} and~\ref{fig:angdist}
makes detailed interpretation premature.  However, we may already note
that the energy dependence appears to have some structure worth
investigating, and that the angular dependence of the $\gamma p\to
K^+K^+\Xi^-$ reaction is different in the two energy ranges covered by
the $g6a$ ($3.2-3.9$ GeV) and $g6b$ ($3.0-5.2$ GeV) data sets.

\subsection{$\Xi^0$ photoproduction}
\begin{figure}[htb]
\includegraphics[width=75mm]{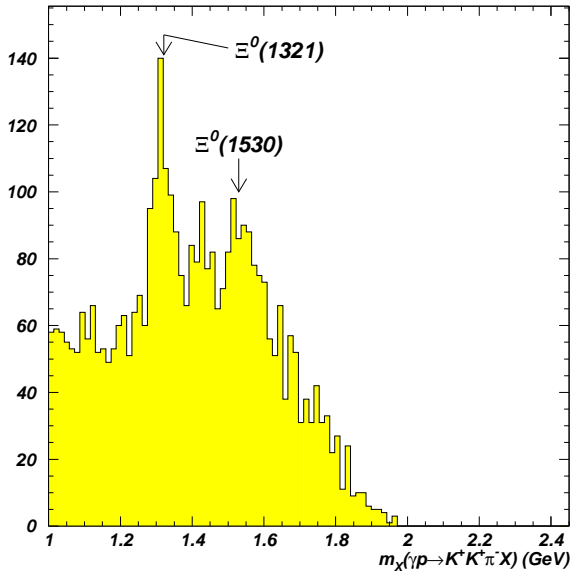}
\caption{\label{fig:g6c-xi0}The $K^+K^+\pi^-$ missing mass in the
  process $\gamma p\to K^+K^+\pi^-X$ from the $g6c$ data set.  There
  are two peaks in this plot that correspond to the known
  $\Xi^0(1321)$ ground state and the first excited state at 1530 MeV.
  The structure between these two peaks is believed to be a $\pi/K$
  misidentification reflection due to the $K^+\pi^+\pi^-\Lambda(1520)$
  final state, and is currently under study.}
\end{figure}

We may look for the $\Xi^0$ in the $K^+K^+\pi^-$ missing mass in the
process $\gamma p\to K^+K^+\pi^-X$.  Detecting the $\pi^-$ complicates
the analysis; because of the toroidal geometry of CLAS, the acceptance
for positive and negative particles is very different.  If the $K^+$
are detected with high acceptance, the $\pi^-$ acceptance is
correspondingly small.  The $g6c$ data set was taken with a reduced
magnetic field and an upstream target position, both of which improved
the acceptance for negatively charged particles.  In this data set, we
can look for the $\Xi^0$.  Fig.~\ref{fig:g6c-xi0} shows our first
result for this search.
This analysis is still underway.  The preliminary result is that we
observe the $\Xi^0(1321)$ and the $\Xi^0(1530)$, with a third peak
between them at around 1410 MeV.  This peak is thought to be a $\pi/K$
misidentification reflection due to the process $\gamma p\to
K^+\pi^+\pi^-\Lambda(1520)$.

\subsection{$\Xi^{--}_5$ pentaquark search}
The recent discovery of the $\Theta^+$ pentaquark and its confirmation
by many groups around the world using different techniques has
generated a great deal of excitement in the nuclear physics community.
The $\Theta^+$ is predicted to be part of an antidecuplet, of which
two other members are also manifestly exotic, in that they cannot be
composed of three-quark states.  These states are the $\Xi^{--}_5$ and
the $\Xi^+_5$.  Finding these states is tremendously important to our
understanding of the nature of the pentaquarks.  Several different
models exist to describe their structure, and a determination of even
just the mass of these states will help to constrain the models.

\begin{figure}[htb]
\begin{minipage}[t]{75mm}
\includegraphics[width=75mm]{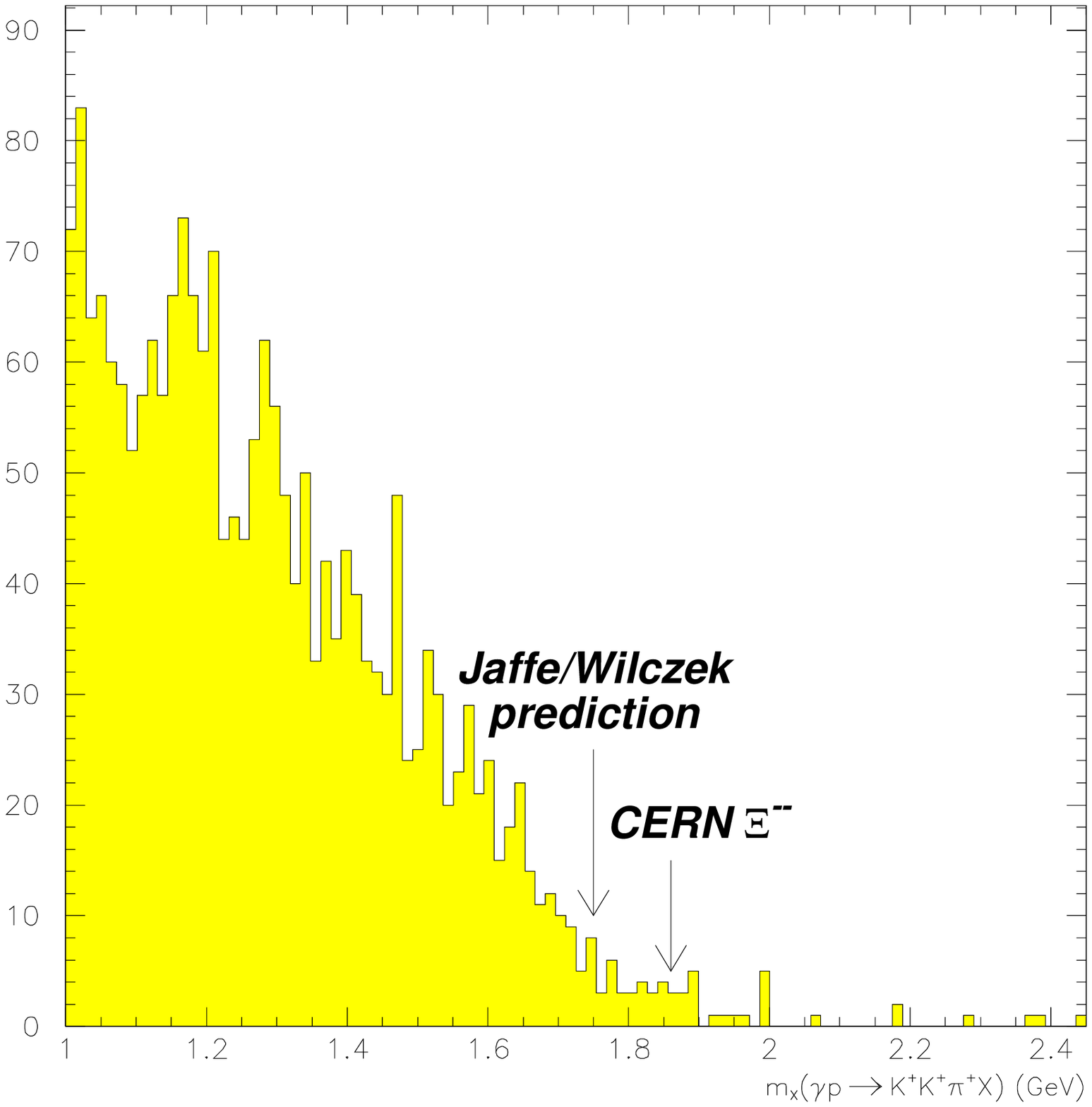}
\caption{\label{fig:g6b-xi}The $K^+K^+\pi^+$ missing mass for the
  $g6b$ data set.  Arrows mark both the initial prediction of Jaffe
  and Wilczek at 1750 MeV, and the position of the NA49 peak at 1860
  MeV.  The phase space in this plot dies out too early to see any
  significant structure.}
\end{minipage}
\hspace{\fill}
\begin{minipage}[t]{75mm}
\includegraphics[width=75mm]{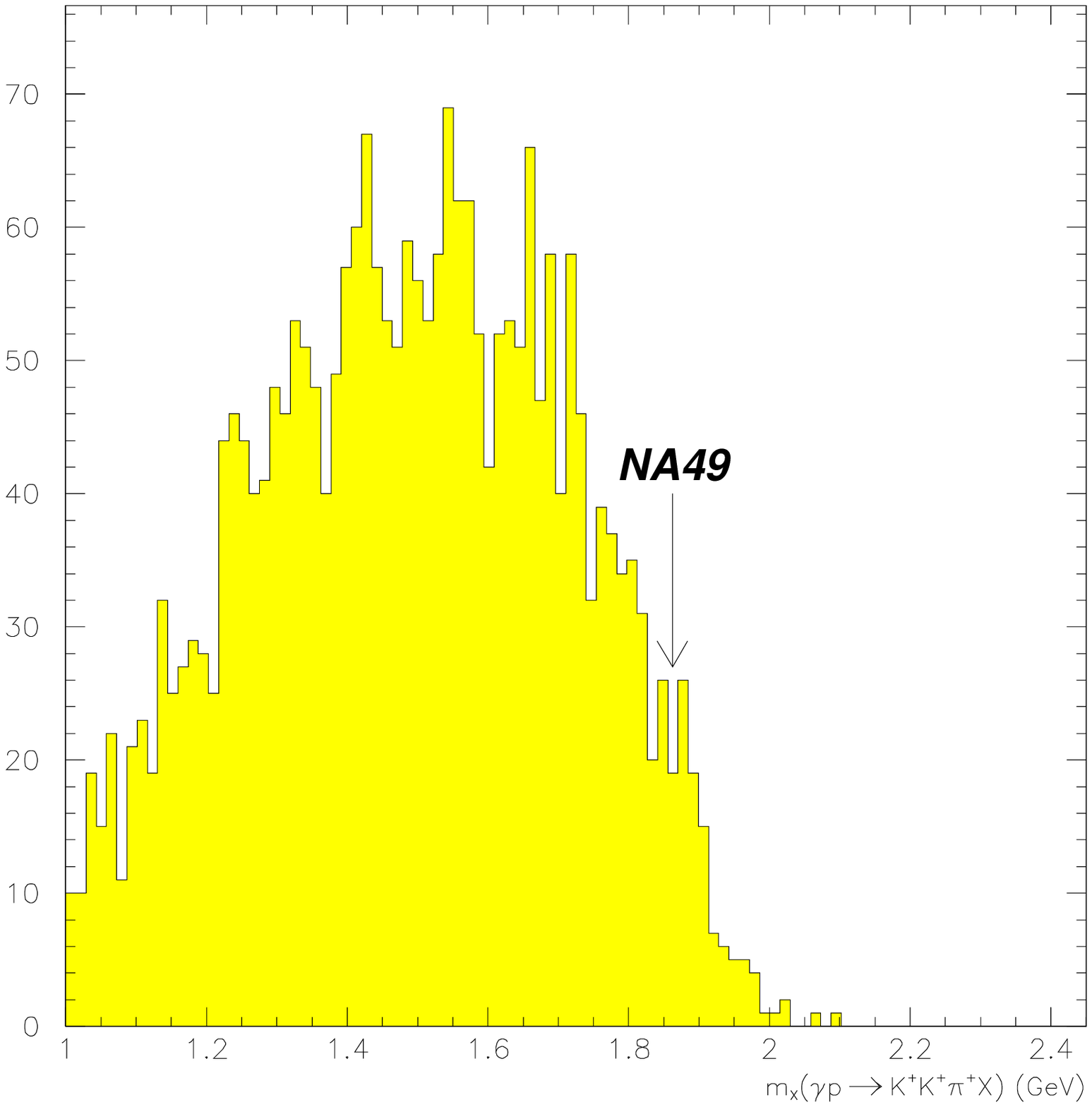}
\caption{\label{fig:g6c-xi}The $K^+K^+\pi^+$ missing mass plot for the
  $g6c$ data set.  An arrow marks the location of the NA49 peak at
  1860 MeV.  No significant structure is seen in the plot.}
\end{minipage}
\end{figure}

The NA49 group has recently claimed to have seen the $\Xi^{--}_5$ in
$pp$ collisions at 17.2~GeV~\cite{Alt03}.  It is vitally important to
confirm or refute this discovery as soon as possible.  We can look for
this state in the $g6b$ and $g6c$ data sets by looking for a peak in
the $K^+K^+\pi^+$ missing mass in the process $\gamma p\to
K^+K^+\pi^+X$.  Figure~\ref{fig:g6b-xi} shows this missing mass for
the $g6b$ data set, and Fig.~\ref{fig:g6c-xi} shows the missing mass
plot for the $g6c$ data set.
No significant structure is seen in either of the plots.  A simple
estimate indicates that there should only be approximately $5-10$
$\Xi^{--}_5$ events in Fig.~\ref{fig:g6c-xi}.

\section{Conclusions}
The interest in cascade physics has increased greatly over the past
year.  We have data that appears to agree with several states listed
in the Particle Data Book, and potentially new structure at 1770 and
1860 MeV.  The structure at 1860 MeV is particularly interesting, as
it corresponds well to the NA49 discovery of the $\Xi^{--}_5$
pentaquark, a state whose properties must be determined in order to
understand the pentaquark structure.  The CLAS detector at Jefferson
Lab is an excellent facility for the study of these states.  The
existing cascade physics program at JLab is well-placed to pursue this
study.  There are currently three separate experimental proposals to
search for the $\Xi^{--}_5$, in an attempt to confirm or refute the
NA49 discovery, and Jefferson Lab will produce some of the best new
information on $\Xi_5$ states in the next few years.


\begin{thebibliography}{9}
\bibitem{Nef95} B.M.K.\ Nefkens, in \textit{Baryons '95}, edited by
  B.\ Gibson \textit{et al.} (World Scientific, Singapore, 1995), 177. 
\bibitem{PDG02} K. Hagiwara \emph{et al.}, Phys.\ Rev.\ D \textbf{66},
  010001 (2002).
\bibitem{Nef96} B.M.K.\ Nefkens, in \textit{$N^{*}$ Physics}, edited
  by T.-S. H. Lee and W. Roberts (World Scientific, Singapore, 1996),
  186. 
\bibitem{Sob00} D. Sober \textit{et al.}, Nucl.\ Instrum.\ Methods
  \textbf{A440}, 263 (2000).
\bibitem{Mec03} B.A. Mecking \emph{et al.}, Nucl.\ Instrum.\ Methods
  \textbf{A503}, 513 (2003). 
\bibitem{Cox00} Aubrey Cox, UCLA note UN-101.
\bibitem{Ris01} D.-O. Riska, in \emph{Proceedings of the Workshop on
the Physics of Excited Nucleons (NSTAR 2001), Mainz, Germany, 2001},
edited by D. Drechsel and L. Tiator (World Scientific, 2001), 129.
\bibitem{Alt03} C. Alt \emph{et al.}, arXiv:hep-ex/0310014.
\end{thebibliography}
\end{document}